\documentclass[11pt]{article}

\usepackage{hyperref,bm,amsmath,amssymb,a4wide,graphicx,cite,subfigure}

\begin{document}

\title{\textbf{Generation of the relic neutrino asymmetry in a hot plasma of the early Universe}}

\author{Victor B. Semikoz$^{a}$\thanks{semikoz@yandex.ru} , Maxim Dvornikov$^{a,b}$\thanks{maxdvo@izmiran.ru} 
\\
$^{a}$\small{\ Pushkov Institute of Terrestrial Magnetism, Ionosphere} \\
\small{and Radiowave Propagation (IZMIRAN),} \\
\small{108840 Troitsk, Moscow, Russia;} \\
$^{b}$\small{\ Physics Faculty, National Research Tomsk State University,} \\
\small{36 Lenin Avenue, 634050 Tomsk, Russia}} 

\date{}

\maketitle

\begin{abstract}
The neutrino asymmetry in the early universe plasma, $n_\nu - n_{\bar \nu}$, is calculated both before and after the electroweak phase transition (EWPT). In the Standard Model before EWPT, the leptogenesis is well known to be driven by the abelian anomaly in a massless hypercharge field. The generation of the neutrino asymmetry in the Higgs phase after EWPT, in its turn, has not been considered previously because of the absence of any quantum anomaly in an external electromagnetic field for such electroneutral particles as neutrino, unlike the Adler-Bell-Jackiw anomaly for charged left and right polarized massless electrons in the same electromagnetic field. Using the neutrino Boltzmann equation, modified by the Berry curvature term in the momentum space, we establish the violation of the macroscopic neutrino current in plasma after EWPT and exactly reproduce the nonconservation of the lepton current in the symmetric phase before EWPT arising in quantum field theory due to the nonzero lepton hypercharge and corresponding triangle anomaly in an external hypercharge field. In the last case, the non-conservation of the lepton current is derived through the kinetic approach without a computation of corresponding Feynman diagrams. Then the new kinetic equation is applied for the calculation of the neutrino asymmetry accounting for the Berry curvature and the electroweak interaction with background fermions in the Higgs phase. Such an interaction generates a neutrino asymmetry through the electroweak coupling of neutrino currents with electromagnetic fields in plasma, which is $\sim G_\mathrm{F}^2$. It turns out that this effect is especially efficient for maximally helical magnetic fields. 
\end{abstract}

\section{Introduction}

There is a common opinion that lepto- and baryo-geneses are originated by the anomalies which are either abelian or non-abelian, or both, and preserve the 't~Hooft's relations for baryon and lepton numbers, $B/3 - L_a=\text{const}$, $a=e,\mu, \tau$. As one possibility to generate a baryon asymmetry of the universe (BAU) through the leptogenesis before the electroweak phase transition (EWPT) at temperatures $T\geq T_\mathrm{EWPT} \simeq 100\,{\rm GeV}$, recently in Refs.~\cite{Semikoz:2013xkc,Dvornikov:2012rk} we reconsidered the well-known baryogenesis scenario with an initial  right electron asymmetry~\cite{Campbell:1992jd} obeying the triangle abelian anomaly in the massless hypercharge field $Y_{\mu}$~\cite{Giovannini:1997eg}. We elucidated in Refs.~\cite{Semikoz:2013xkc,Dvornikov:2012rk} that the similar abelian anomaly for the left lepton  doublet $L=(\nu_{e\mathrm{L}},e_\mathrm{L})^\mathrm{T}$,
\begin{equation}\label{abelian}
  \frac{\partial j^{\mu}_\mathrm{L}}{\partial x^{\mu}} =
  -\frac {g'^{2}}{16\pi^2}
  ({\bf E}_\mathrm{Y}\cdot{\bf B}_\mathrm{Y}), 
\end{equation}
where $g'=e/\cos \theta_W$ is the Standard Model (SM) coupling, $\sin^2\theta_W=0.23$ is the Weinberg parameter, and ${\bf E}_\mathrm{Y}= -\partial_t{\bf Y} - \nabla Y_0$, ${\bf B}_\mathrm{Y}= \nabla\times {\bf Y}$ are the hyperelectric and hypermagnetic fields correspondingly, plays rather a secondary role in baryogenesis for the chosen initial conditions since a left lepton asymmetry does not have enough time to be developed before EWPT in such a way to wash out BAU through sphalerons which interact with left constituents of primordial plasma only. 

In any way, one can expect that after EWPT there appears at $T<T_\mathrm{EWPT}$ an initial neutrino asymmetry $n_{\nu_a} - n_{\bar{\nu}_a}=\xi_{\nu_a}T^3/6$, where $\xi_{\nu_a}=\mu_{\nu_a}/T\neq 0$ is the asymmetry parameter and $\mu_{\nu_a}$ is the neutrino chemical potential~\cite{fn1}. The concrete initial neutrino asymmetry in a hot plasma at $\mathcal{O}({\rm MeV})< T < T_\mathrm{EWPT}$  is unknown except of the primordial nucleosynthesis  (upper)  bound on it, $|\xi_{\nu_a}| < 0.07$~ at $T_\mathrm{BBN}=0.1\,{\rm MeV}$ \cite{Dolgov:2002ab}, accounting for neutrino oscillations with equivalent $\xi_{\nu_e}\sim \xi_{\nu_{\mu}}\sim \xi_{\nu_{\tau}}$ at $T\sim \mathcal{O}({\rm MeV})$ \cite{Pastor:2008ti}. Let us try to find other ways to put a bound on relic neutrino asymmetries before the BBN time.

Since a neutrino has the zero electric charge, there is no triangle anomaly for it after EWPT, in contrast to charged fermions in QED. Let us remind that the Adler's triangle anomaly for chiral (massless) charged fermions,
\begin{equation}\label{Adler}
  \partial_t n_\mathrm{R,L} + (\nabla\cdot {\bf j}_\mathrm{R,L}) =
  \pm \frac{\alpha_\mathrm{em}}{\pi} ({\bf E}\cdot{\bf B}),
\end{equation}
provides the instability of a seed magnetic field in a relativistic plasma, e.g., in the case of the hot plasma of early universe in the presence of a difference between the chemical potentials  (densities) of right-handed and left-handed charged particles, $\mu_5=(\mu_\mathrm{R} - \mu_\mathrm{L})/2\neq 0$, which is diminishing because of the helicity flip when the fermion mass is accounted for~\cite{Boyarsky:2011uy}.

Note that, recently, the anomaly in Eq.~\eqref{Adler} was reproduced independently through the Boltzmann kinetic equation accounting for the Berry curvature in the momentum space in Ref.~\cite{Son:2012zy} (see Eq. (20) therein). Such an equation is extensively considered and widely applied in condensed matter physics, as well as in the studies of heavy ion collisions with the formation of quark-gluon plasma~\cite{Kharzeev:2015kna}. The Berry curvature modification of the kinetic theory accounting, in particular, for the chiral magnetic effect in inhomogeneous chiral plasma has been recently done in Ref.~\cite{Gorbar:2016qfh}. Below we consider the chiral matter of massless neutrinos before and after EWPT, described by the kinetic Boltzmann equation, in both  cases accounting for the Berry curvature and obtain the analogues of the anomaly in Eq.~\eqref{Adler} with the nonconservation of the four-current for the $\nu\bar{\nu}$ gas.

Our work is organized as follows. In Sec.~\ref{sec:BERRY}, we recall the concept of the Berry curvature, which is the induced gauge field related to the topological Berry phase~\cite{Berry,Dubovik}, for neutrinos. Then, in Sec.~\ref{RKE}, starting with the Boltzmann equations without such a curvature, we generalize them by adding the terms responsible for this topological effect. This generalization is made in the two situations: (a) in the presence of massless hypercharge fields before EWPT; and mostly (b) in a hot plasma after EWPT at temperatures $T\ll T_\mathrm{EWPT}$ when the Fermi approximation is valid for the neutrino interactions with plasma. In the case~(a) we reproduce  the well-known abelian anomaly for the left current $j_\mathrm{L}^{\mu}$ in Eq.~(\ref{abelian}). In the case~(b), after EWPT,  we predict the new anomalous violation of the neutrino current, $\partial_{\mu}j^{\mu}_{\nu_a}\neq 0$, unknown before in literature. Then, in Sec.~\ref{afterEWPT}, we describe the time evolution of the neutrino asymmetry during the universe expansion using such a new anomaly. We summarize our results in Sec.~\ref{Conclusions}.
In Appendix~\ref{sec:CONTEQ}, we analyze the collision integrals in the neutrino kinetic equations in a primordial plasma.

\section{Berry curvature\label{sec:BERRY}}

We consider massless neutrinos, $m_{\nu_a}=0$,  $a=e,\mu,\tau$, i.e. we neglect neutrino oscillations.
These particles participate in electroweak interactions in SM as the chiral ones, $\psi_{\nu}\equiv \psi_{\nu_\mathrm{L}}=(1 - \gamma^5)\psi_{\nu}/2$, thus neutrinos are left-handed, $(\bm{\sigma}\cdot\mathbf{k})u_-({\bf k})=-ku_-({\bf k})$, while antineutrinos are right-handed, $(\bm{\sigma}\cdot\mathbf{k})u_+({\bf k})=ku_+({\bf k})$. Here
\begin{equation}\label{eq:2cspinors}
  u_-({\bf k}) =
  \begin{pmatrix}
    -e^{-\mathrm{i}\varphi}\sin \theta/2
    \\
    \cos \theta/2 \
  \end{pmatrix},
  \quad
  u_+({\bf k}) =
  \begin{pmatrix}
    e^{-\mathrm{i}\varphi}\cos \theta/2
    \\
    \sin \theta/2 \
  \end{pmatrix},
\end{equation}
are the two-component spinors respectively and $\bm{\sigma}$ are the Pauli matrices. These spinors define a nonzero Berry connection, which are the components of the induced gauge field in the momentum space \cite{Dubovik}:
\begin{equation}\label{connection}
  {\bf a}_{\bf k}^{\pm}=(a_k^{\pm},a_{\varphi}^{\pm},a_{\theta}^{\pm})=   
  \mathrm{i}u_{\pm}^+({\bf k})\nabla_{\bf k}u_{\pm}({\bf k}).
\end{equation}
Using the spinors in Eq.~\eqref{eq:2cspinors}, one can easily calculate the components of this field in spherical coordinates, $a_k^{\pm}=a_{\theta}^{\pm}=0$, $a_{\varphi}^{-}= \tan (\theta/2)/2k$, and
$a_{\varphi}^{+}=\cot (\theta/2)/2k$, that leads to the Berry curvature,
\begin{equation}\label{curvature}
  \bm{\Omega}_{\bf k}^{\pm} =
  \nabla_{\bf k}\times {\bf a}_{\bf k}^{\pm} =
  \pm \frac{\hat{\bf k}}{2k^2},
  \quad
  {\bf n}\equiv\hat{\bf {k}}=\frac{\bf k}{k},
  \quad
  n^2=1,
\end{equation}
where the upper sign stays for $\bar{\nu}_a$ and the lower one for $\nu_a$. The Berry connection in Eq.~(\ref{connection}) enters as the additional term in the action for chiral right-and left-polarized charged particles in the presence of an electromagnetic field $A^{\mu}=(A_0, {\bf A})$~\cite{Son:2012zy}, 
\begin{equation}\label{action}
  S = \int \mathrm{d}t
  \left[
    ({\bf p} - e{\bf A})\dot{\bf x} - (\varepsilon_{\bf p} - eA_0) -
    {\bf a}^{\pm}_{\bf p}\cdot\dot{\bf p}
  \right].
\end{equation}
An analogous relation holds for a lepton (in particular, a left-handed neutrino) interacting with a hypercharge field before an EWPT through the
SM constants $ g_\mathrm{R,L} = g' y_\mathrm{R,L}/2$, where $y_\mathrm{L}=-1$ is the hypercharge for the left doublet and $y_\mathrm{R}=-2$ is the hypercharge for the right-handed singlet (right-handed electron),
\begin{equation}\label{action2}
  S=\int \mathrm{d}t
  \left[
    ({\bf k} - g_\mathrm{R,L}{\bf Y})\dot{\bf x}  -
    (\varepsilon_{\bf k} - g_\mathrm{R,L}Y_0)- {\bf a}^{\pm}_{\bf k}\cdot\dot{\bf k}
  \right].
\end{equation}
In the latter case, we neglect the contribution of the pseudovector current~\cite{fn0}
%
%
in the Lagrangian of the interaction of leptons with a massless hypercharge field,
\begin{equation}
  \mathcal{L}_\mathrm{int} = 
  \sum_a
  [g_\mathrm{L}\bar{L}_a\gamma_{\mu}L_a +   
  g_\mathrm{R}\bar{l}_{a\mathrm{R}}\gamma_{\mu}l_{a\mathrm{R}}]Y^{\mu},
\end{equation}
accounting for  the vector interaction in Eq.~(\ref{action2}) only by analogy with the vector electromagnetic interaction, $\mathcal{L}_\mathrm{em}=e\bar{\psi}\gamma_{\mu}\psi A^{\mu}$. This fact simplifies the derivation of the Boltzmann kinetic equation
\begin{equation}
  \frac{\partial f_\mathrm{R,L}}{\partial t} +
  \dot{\bf x} \frac{\partial f_\mathrm{R,L}}{\partial {\bf x}} +
  \dot{\bf k} \frac{\partial f_\mathrm{R,L}}{\partial {\bf k}} = J_\mathrm{coll},
\end{equation}
see below.

Finally, to derive the kinetic equation from the action for a neutrino in an unpolarized medium after EWPT, in the Fermi approximation at temperatures $T\ll T_\mathrm{EWPT}$,
\begin{equation}\label{action3}
  S=\int \mathrm{d}t
  \left[
    ({\bf k} - G_\mathrm{F}\sqrt{2}c^a_\mathrm{V}\delta {\bf j}^{(e)})\dot{\bf x}  - 
    (\varepsilon_{\bf k} - G_\mathrm{F}\sqrt{2}c^a_\mathrm{V}\delta n^{(e)} )-
    {\bf a}^{\pm}_{\bf k}\cdot\dot{\bf k}
  \right],
\end{equation}
we write down the corresponding equations of motion. Taking into account the Berry curvature in Eq.~(\ref{curvature}), these equations have the form (see the case of charged particles in Refs.~\cite{Son:2012zy,Yamamoto:2015gzz}):
\begin{eqnarray}\label{motion}
  &&\dot{\bf x}=\frac{\partial \varepsilon_{\bf k}}{\partial {\bf k}} -
  (\dot{\bf k}\times {\bf \bm{\Omega}_{\bf k}^{\pm}}),
  \nonumber
  \\&&
  \dot{\bf k}=G_\mathrm{F}\sqrt{2}c_\mathrm{V}^a
  \left[
    -\frac{\partial \delta {\bf j}^{(e)}}{\partial t} -\nabla \delta n^{(e)} +
    \dot {\bf x}\times (\nabla\times \delta {\bf j}^{(e)}) 
  \right].   
\end{eqnarray}
For the nonlinear $\nu\nu$ interactions, one should substitute the coefficient $2G_\mathrm{F}\sqrt{2}$ instead of $G_\mathrm{F}\sqrt{2}c_\mathrm{V}^{(a)}$ with the change of the superscript $e\to \nu$ for the number (and three-current) densities.
Here $G_\mathrm{F}=1.17\times10^{-5}\thinspace\text{GeV}^{-2}$ is the Fermi constant, $c_\mathrm{V}^{(a)}=2\xi \pm 0.5$ is the vector coupling constant for
$\nu_a e$ interactions (upper sign stays for electron neutrinos), $\xi=\sin^2\theta_\mathrm{W}=0.23$ is the Weinberg parameter in SM, $\delta n^{(e)}({\bf x},t)=n_e({\bf x},t) - n_{\bar{e}}({\bf x},t)$ is the asymmetry of the electron number density in $e^-e^+$ plasma, and $\delta {\bf j}^{(e)}({\bf x},t)={\bf j}_e({\bf x},t) - {\bf j}_{\bar{e}}({\bf x},t)$ is the asymmetry of the electron three-current density. We note the complexity of the coupled equations of motion in Eq.~\eqref{motion} when the Berry curvature is taken into account, when, after some algebraic transformations, while decoupling the velocities $\dot{\bf x}$ and the forces $\dot{\bf x}$, their expressions and the phase volume $\mathrm{d}^3 x \mathrm{d}^3 k$ change, see below.

\section{Neutrino kinetic equations accounting for the Berry curvature\label{RKE}}

\paragraph{(a) Without a Berry curvature.} It is not surprising that, if one takes into account only vector interaction with a hypercharge field without a Berry curvature, in full analogy with a usual Boltzmann kinetic equation for charged particles, the kinetic equation for neutrinos and antineutrinos in the symmetric phase of the early universe has the form,
\begin{multline}\label{RKEbefore}
  \frac{\partial f^{(\nu_a,\bar{\nu}_a)}({\bf k},{\bf x},t)}{\partial t} +
  {\bf n}\frac{\partial f^{(\nu_a,\bar{\nu}_a)}({\bf k},{\bf x},t)}{\partial {\bf x}}
  \\
  \pm g_\mathrm{L}
  \Bigl[
    {\bf E}_\mathrm{Y}({\bf x},t) +
    {\bf n}\times {\bf B}_\mathrm{Y}({\bf x},t)
  \Bigr]
  \frac{\partial f^{(\nu_a,\bar{\nu}_a)}({\bf k},{\bf x},t)}{\partial {\bf k}}=
  J^{(\nu_a,\bar{\nu}_a)}({\bf k},{\bf x},t),
\end{multline}
where $J^{(\nu_a,\bar{\nu}_a)}$ are the collision integrals and $f^{(\nu_a,\bar{\nu}_a)}({\bf k},{\bf x},t)$ are the distribution functions for neutrinos (antineutrinos)  with the upper (lower) sign in the force term correspondingly.

Let us remind that the Boltzmann equation for neutrinos (antineutrinos) in unpolarized matter at the temperature $T\ll T_\mathrm{EWPT}$ and also without a Berry curvature has the form~\cite{Silva:1999zz,Oraevsky:2001gd,Bento:1999tj,Semikoz:2004bq}, which results from the action in Eq.~\eqref{action3}, when for $\bm{\Omega}_\mathbf{k} = 0$, the velocity becomes a usual unit velocity of a massless particle, $\dot{\mathbf{x}} = \partial \varepsilon_\mathbf{k} / \partial \mathbf{k} = \mathbf{n}$:
\begin{multline}\label{Boltzmann} 
  \frac{\partial f^{(\nu_a,\bar{\nu}_a)}({\bf k},{\bf x},t)}{\partial t} +
  {\bf n}\frac{\partial f^{(\nu_a,\bar{\nu}_a)}({\bf k},{\bf x},t)}{\partial {\bf x}}
  \\
  \pm
  \Bigl[
    {\bf E}_e({\bf x},t) + {\bf n}\times {\bf B}_e({\bf x},t)
  \Bigr]
  \frac{\partial f^{(\nu_a,\bar{\nu}_a)}({\bf k},{\bf x},t)}{\partial {\bf k}}=
  J^{(\nu_a,\bar{\nu}_a)}({\bf k},{\bf x},t),
\end{multline}
where for massless $\nu_a$ ($\bar{\nu}_a$) one substitutes the upper (lower) sign for the third (force) term given by the weak $\nu e$ interactions in the Fermi approximation:
\begin{eqnarray}\label{force}
  &&{\bf E}_e({\bf x},t) =
  G_\mathrm{F}\sqrt{2}c_\mathrm{V}^a
  \Bigl[
    - \nabla \delta n^{(e)}({\bf x},t) -\frac{\partial \delta{\bf j}^{(e)}({\bf x},t)}{\partial t}
  \Bigr],
  \nonumber
  \\
  && {\bf B}_e({\bf x},t)=G_\mathrm{F}\sqrt{2}c_\mathrm{V}^a
  \nabla\times \delta{\bf j}^{(e)}({\bf x},t).
\end{eqnarray}
It is interesting to mention that, since the force term in Eq.~(\ref{Boltzmann}) has the Lorentz form, the corresponding effective electromagnetic fields in Eq.~(\ref{force}) obey the standard Maxwell equations: $(\nabla\cdot {\bf B}_e)=0$ and $\partial_t{\bf B}_e= -(\nabla \times {\bf E}_e)$.

Let us stress that, neglecting the Berry curvature, we have the spectrum $\varepsilon_{\bf k}=k$, for which the four-current
\begin{equation}
  j_{\mu}^{(\nu_a,\bar{\nu}_a)}({\bf x},t) =
  (n_{\nu_a, \bar{\nu}_a}({\bf x},t), {\bf j}_{\nu_a, \bar{\nu}_a}({\bf x},t)) =
  \int \frac{\mathrm{d}^3 k}{(2\pi)^3}
  \frac{k_{\mu}}{\varepsilon_k}
  f^{(\nu_a,\bar{\nu}_a)}({\bf k},{\bf x},t),
\end{equation}
is conserved,
as results from both kinetic Eqs.~(\ref{RKEbefore}) and~(\ref{Boltzmann}) integrated over $\mathrm{d}^3k$:
\begin{equation}\label{current}
  \frac{\partial j_{\mu}^{(\nu_a,\bar{\nu}_a)}({\bf x},t)}{\partial x_{\mu}} =
  \frac{\partial n^{(\nu_a,\bar{\nu}_a)}({\bf x},t)}{\partial t} +
  \frac{\partial [{\bf V}^{(\nu_a,\bar{\nu}_a)}({\bf x},t) \cdot
  n^{(\nu_a,\bar{\nu}_a)}({\bf x},t)]}{\partial {\bf x}}=0,
\end{equation}
where the macroscopic neutrino fluid velocity of the $\nu_a\bar{\nu}_a$ gas
\begin{equation}
  {\bf V}^{(\nu_a,\bar{\nu}_a)}({\bf x},t) =
  \frac{1}{n^{(\nu_a,\bar{\nu}_a)}({\bf x},t)}
  \int \frac{\mathrm{d}^3k}{(2\pi)^3}{\bf n}
  f^{(\nu_a,\bar{\nu}_a)}({\bf k},{\bf x},t),
\end{equation}
can be non-relativistic, $|{\bf V}|\ll 1$, contrary to the microscopic one, $|{\bf n}|=1$. In Appendix~\ref{sec:CONTEQ} we demonstrate that the collision integral, being integrated over the neutrino momentum, does not contribute to the conservation law in Eq.~(\ref{current}), $\smallint \mathrm{d}^3 k J^{(\nu_a,\bar{\nu}_a)}({\bf k},{\bf x},t)=0$, both for elastic and inelastic neutrino (antineutrino) scattering.

\paragraph{(b) Accounting for the Berry curvature.} Now let us turn to the case of the Berry curvature in Eq.~(\ref{curvature}) considering, e.g., a generalization of the neutrino kinetic Eq.~(\ref{Boltzmann}).
In full analogy with the approach in Refs.~\cite{Son:2012zy,Yamamoto:2015gzz}, one can write for the chiral fermions, having the modified
dispersion relation,
\begin{equation}\label{spectrum}
  \varepsilon_{\bf k}=k[1 - \bm{\Omega}_{\bf k}\cdot\mathbf{B}_e({\bf x},t)],\end{equation}
the modified  Boltzmann equation for the neutrino distribution function $f_{\bf k}^{(\nu_a)}\equiv f^{(\nu_a)}({\bf k},{\bf x},t)$ (for simplicity, for neutrinos only),
\begin{multline}\label{Boltzmann2}
  \frac{\partial f_{\bf k}^{(\nu_a)}}{\partial t} +
  \frac{1}{\sqrt{\omega}}\left(\tilde{\bf v} +
  \tilde{\bf E}_e\times \bm{\Omega}_{\bf k} +
  (\tilde{\bf v}\cdot\bm{\Omega}_{\bf k}){\bf B}_e\right)
  \frac{\partial f_{\bf k}^{(\nu_a)}}{\partial {\bf x}}
  \\
  +\frac{1}{\sqrt{\omega}}\left(\tilde{\bf E}_e +
  \tilde{\bf v}\times {\bf B}_e +
  (\tilde{\bf E}\cdot{\bf B}_e)\bm{\Omega}_{\bf k}\right)
  \frac{\partial f_{\bf k}^{(\nu_a)}}{\partial {\bf k}} =
  J^{(\nu_a)}(f_{\bf k}^{(\nu_a)}),
\end{multline}
where $\omega=(1 + {\bf B}_e\cdot\bm{\Omega}_{\bf k})^2$ is the factor, defining the invariant phase space, $\mathrm{d}^3 k \mathrm{d}^3 x \to \sqrt{\omega}\mathrm{d}^3 k \mathrm{d}^3 x$, which is given by the Berry curvature in Eq.~(\ref{curvature}) and the effective magnetic field in Eq.~(\ref{force}); $\tilde{\bf v}=\partial \varepsilon_{\bf k}/\partial {\bf k}$ is the effective neutrino velocity, and $\tilde{\bf E}_e={\bf E}_e - \partial \varepsilon_{\bf k}/\partial {\bf x}$ is the effective electric field in the modified Boltzmann Eq.~(\ref{force}), both given by the spectrum in Eq.~(\ref{spectrum}).

The neutrino number density and the neutrino three-current density,
\begin{align}
  \label{density}
  n^{(\nu_a)}({\bf x},t) = &
  \int \frac{\mathrm{d}^3k}{(2\pi)^3}\sqrt{\omega}f_{\bf k}^{(\nu_a)},
  \\
  \label{3current}
  {\bf j}^{(\nu_a)}({\bf x},t) = & 
  \int \frac{\mathrm{d}^3k}{(2\pi)^3}\left(\tilde{\bf v} + \tilde{\bf E}_e\times \bm{\Omega}_{\bf k} + (\tilde{\bf v}\cdot\bm{\Omega}_{\bf k}){\bf B}_e\right)f_{\bf k}^{(\nu_a)},
\end{align}
obey the quantum anomaly (the non-conservation of the neutrino four-current, $\partial^{\mu}j_{\mu}^{(\nu_a)}\neq 0$) due to the weak interactions in SM in the presence of the Berry curvature (compare with Ref.~\cite{Son:2012zy}),
\begin{align}\label{anomaly}
  \partial_tn^{(\nu_a)} + \nabla\cdot{\bf j}^{(\nu_a)} = &
  - ({\bf E}_e\cdot {\bf B}_e)
  \int \frac{\mathrm{d}^3k}{(2\pi)^3}
  \left(
    \bm{\Omega}_{\bf k}\cdot\frac{\partial f^{(\nu_a)}_{\bf k}}{\partial {\bf k}}
  \right)
  \nonumber
  \\
  & =
  - C^{(\nu_a)}({\bf E}_e\cdot {\bf B}_e)\neq 0.
\end{align}
One can easily calculate the integral in Eq. (\ref{anomaly}) for neutrinos with the Fermi distribution
\begin{equation*} 
  f^{(\nu_a)}(k)=\frac{1}{\exp [(k - \mu_{\nu_a})/T] + 1},
\end{equation*}
substituting $\Omega_{\bf k}= - {\bf k}/2k^3$ from Eq. (\ref{curvature}) for neutrinos~\cite{fn2},
\begin{equation}\label{integral0}
  C^{(\nu_a)} = \frac{1}{4\pi^2T}
  \int_0^{\infty}\mathrm{d}k
  \frac{e^{(k - \mu_{\nu_a})/T}}{[e^{(k - \mu_{\nu_a})/T} +1]^2} =
  \frac{1}{4\pi^2(1 + e^{-\mu_{\nu_a}/T})}.
\end{equation}
Since for antineutrinos this parameter has opposite sign due to the different sign of the Berry curvature $\bm{\Omega}_{\bf k}$, and accounting for the opposite sign of the chemical potential for $\bar{\nu}_a$, $\mu_{\nu_a}\to - \mu_{\nu_a}$, one can obtain from Eq.~(\ref{anomaly}) the main result of this work for the neutrino asymmetry evolution,
\begin{equation}\label{main}
  \frac{{\rm d}}{{\rm d}t}(n_{\nu_a} - n_{\bar{\nu}_a})=
  -\frac{1}{4\pi^2}\int\frac{\mathrm{d}^3x}{V}({\bf E}_e\cdot{\bf B}_e),
\end{equation}
where the effective electromagnetic fields ${\bf E}_e, {\bf B}_e$ are given by Eq.~(\ref{force}). 

\subsection{Anomalies for lepton currents in the symmetric phase of the early universe\label{sec:ASYMEE}}

Now we are able to obtain the quantum effect of the nonconservation of lepton currents in a hypercharge field $Y^\mu$ from the Boltzmann equation accounting for the Berry curvature. Similarly to the spectrum in Eq.~(\ref{spectrum}) and the neutrino anomaly after EWPT in Eq.~(\ref{main}), introducing the neutrino spectrum $\varepsilon_{\bf k}=k[1 - g_\mathrm{L}\bm{\Omega}_{\bf k}\cdot \mathbf{B}_\mathrm{Y}]$, from the Boltzmann Eq.~\eqref{RKEbefore}, analogously to the modification in Eq.~(\ref{Boltzmann}), which results in Eq.~\eqref{Boltzmann2} and then to the four-current in Eqs.~(\ref{density}),~(\ref{3current}), we derive  the similar anomaly for the left lepton current before EWPT:
\begin{equation}\label{main2}
  \frac{{\rm d}}{{\rm d}t}(n_{\nu_a} - n_{\bar{\nu}_a})=
  -\frac{g_\mathrm{L}^2}{4\pi^2}
  \int\frac{\mathrm{d}^3x}{V}({\bf E}_\mathrm{Y}\cdot{\bf B}_\mathrm{Y})=
  -\frac{g'^{2}}{16\pi^2}
  \int\frac{\mathrm{d}^3x}{V}({\bf E}_\mathrm{Y}\cdot{\bf B}_\mathrm{Y}).
\end{equation}
It is obvious that absolutely the same abelian anomaly exists for left electrons $e_\mathrm{L}$~\cite{fn3}, or their asymmetry density $(n_{e} - n_{\bar{e}})$ obeys Eq.~(\ref{main2}) due to the same initial form of the Boltzman Eq.~(\ref{RKEbefore}).

However, in the case of the right electron $e_\mathrm{R}$ we, firstly, should substitute $g_\mathrm{L}\to g_\mathrm{R}$ in Eq.~(\ref{RKEbefore}), and, secondly, take into account that a massless $e_\mathrm{R}$ is right-handed like an antineutrino, or for it the Berry curvature $\bm{\Omega}_{\bf k}$ leads to the opposite sign in Eq. (\ref{integral0}):
\begin{equation}\label{integral2}
  C^{(e_\mathrm{R})}=
  -\frac{g_\mathrm{R}^2}{4\pi^2T}
  \int_0^{\infty}\mathrm{d}k
  \frac{e^{(k - \mu_{e\mathrm{R}})/T}}{[e^{(k - \mu_{e\mathrm{R}})/T} +1]^2}= 
  -\frac{g_\mathrm{R}^2}{4\pi^2(1 + e^{-\mu_{e\mathrm{R}}/T})}.
\end{equation}
Vice versa, massless right positrons are left-handed like $\nu_{e\mathrm{L}}$ and $e_\mathrm{L}$, or for them one obtains the same sign as in Eq.~(\ref{integral0}) with the change of chemical potential $\mu_{e_\mathrm{R}}\to - \mu_{e_\mathrm{R}}$. Hence
\begin{equation*}
  C^{(\bar{e}_\mathrm{R})}= \frac{g_\mathrm{R}^2}{4\pi^2}
  \frac{1}{1 + \exp(\mu_{e_\mathrm{R}}/T)}.
\end{equation*}
Finally it is not difficult to restore the abelian anomaly for right electrons with the effective charge 
$g_\mathrm{R}=g'y_\mathrm{R}/2$:
\begin{equation}\label{main3}
  \frac{{\rm d}(n_{e_\mathrm{R}} - n_{\bar{e}_\mathrm{R}})}{{\rm d}t} =
  \frac{g_\mathrm{R}^2}{4\pi^2}\int\frac{\mathrm{d}^3x}{V}
  ({\bf E}_\mathrm{Y}\cdot{\bf B}_\mathrm{Y}) =
  \frac{g'^{2}}{4\pi^2}
  \int\frac{\mathrm{d}^3x}{V}
  ({\bf E}_\mathrm{Y}\cdot{\bf B}_\mathrm{Y}).
\end{equation}
Thus, we have just recovered the well-known abelian anomalies for lepton currents in hypercharge fields, not using Feynman diagram techniques.

At the end of this section, we emphasize that, in order to obtain an analog of the abelian anomaly for a neutrino in the Higgs (broken) phase after EWPT as in Eq.~\eqref{main}, we had to consider the electroweak interaction of a neutrino with plasma using the effective electromagnetic fields in Eq.~\eqref{force} and the Berry curvature in Eq.~\eqref{anomaly}.

\section{Neutrino asymmetry generation in a hot plasma}\label{afterEWPT}

In this section we shall apply the neutrino anomaly in Eq.~(\ref{main}) to study the neutrino asymmetry generation in the early Universe. Let us consider the hot plasma at relativistic temperatures much below the EWPT temperature, $m_e\ll T\ll T_\mathrm{EWPT}$, when we can use the Fermi approximation for the electroweak neutrino interactions with matter. Note that, for simplicity, our calculations are limited by the neutrino interaction in the lepton plasma consisting of $e^{+}e^-$. Nevertheless other plasma components above $T>T_\mathrm{QCD}\simeq  100\,{\rm MeV}$ can be included straightforwardly.

The effective electromagnetic fields  ${\bf E}_{e}$ and ${\bf B}_{e}$ are given by Eq.~(\ref{force}). Accounting for the standard Maxwell equations for the usual electromagnetic fields ${\bf E}$ and ${\bf B}$ in the MHD approximation,
\begin{equation}
  {\bf j}_\mathrm{em} = - e\delta{\bf j}^{(e)}= (\nabla\times {\bf B})
  \quad
  \dot{\bf B}= - (\nabla\times {\bf E})
  \quad
  \nabla\cdot{\bf E}=-e\delta n_e
  \quad
  (\nabla\cdot {\bf B})=0,
\end{equation}
where the asymmetries of the number density and of the three-current, $\delta n^{(e)} = n_e - n_{\bar e}$ and $\delta \mathbf{j}^{(e)} = \mathbf{j}_e - \mathbf{j}_{\bar e}$, are defined below Eq.~\eqref{motion}, we get that the effective fields in Eq.~\eqref{force}, which arise owing to the electroweak interaction, can be expressed directly through the Maxwell fields:
\begin{equation}
  {\bf E}_{e}({\bf x},t)= A\nabla^2{\bf E}({\bf x},t),
  \quad
  {\bf B}_{e}({\bf x},t)= A\nabla^2{\bf B}({\bf x},t),
\end{equation}
where $A=G_\mathrm{F}\sqrt{2}c_V^a/e$, $e=\sqrt{4\pi\alpha_\mathrm{em}} \sim 0.3 > 0$ is the absolute value of the electron charge, and $\alpha_\mathrm{em} \approx 1/137$ is the fine structure constant.

Then we use the Fourier representation of the electromagnetic field,
\begin{equation}
  {\bf E}({\bf x},t) =
  \int \frac{\mathrm{d}^3k}{(2\pi)^3} e^{\mathrm{i}{\bf k x}}{\bf E}_k(t),
  \quad
  {\bf B}({\bf x},t) =
  \int \frac{\mathrm{d}^3k}{(2\pi)^3} e^{\mathrm{i}{\bf k x}}{\bf B}_{k}(t),
\end{equation}
where one should take into account that ${\bf B}({\bf x},t)={\bf B}^*({\bf x},t)$. In this case, the neutrino asymmetry evolution in Eq.~(\ref{main}),
\begin{align}\label{Fourier}
  \frac{{\rm d}(n_{\nu_a} - n_{\bar{\nu}_a})}{{\rm d}t} = &
  - \frac{A^2}{8\pi^2V}\int \frac{\mathrm{d}^3k}{(2\pi)^3}k^4
  [{\bf E}_k(t)\cdot{\bf B}^*_k(t) + \mathrm{c.c.}] 
  \notag
  \\
  & =
  \frac{A^2}{8\pi^2}\int k^4\frac{\partial}{\partial t}h(k,t)\mathrm{d}k,
\end{align}
is given by an isotropic  spectrum $h(k,t)$ of the magnetic helicity density $h(t)=\smallint \mathrm{d} k h(k,t)$ where
$h(k,t)=k^2[{\bf A}_k(t)\cdot{\bf B}_k^*(t) + \mathrm{c.c.}]/4\pi^2V$, and
\begin{equation}
  \frac{{\rm d}}{{\rm d}t}h(k,t) =
  -\frac{k^2}{2\pi^2V}
  \left[
    {\bf E}_k(t)\cdot{\bf B}_k^*(t) + \mathrm{c.c.}
  \right].
\end{equation}
Substituting the neutrino density asymmetry $n_{\nu_a} - n_{\bar{\nu}_a}=T^3\xi_{\nu_a}(T)/6$, defining the variable $\xi_{\nu_a}=\mu_{\nu_a}(T)/T$, where $\mu_{\nu_a}$ is the neutrino chemical potential, and using the conformal dimensionless variables, $t\to \eta=M_0/T$, $a=T^{-1}$, $\tilde{h}(\tilde{k},\eta)=a^2h(k,t)$, where $\tilde{k} = ak$ is a constant value, one can recast master Eq.~(\ref{Fourier}) in the comoving volume:
\begin{equation}\label{conformal}
  \frac{{\rm d}\xi_{\nu_a}(\eta)}{{\rm d}\eta} =
  \frac{3A^2}{4\pi^2a^2}
  \int \tilde{k}^4 \frac{{\rm d}}{{\rm d}\eta}
  \left[
    \frac{\tilde{h}(\tilde{k},\eta)}{a^2}
  \right] \mathrm{d}\tilde{k}.
\end{equation}

The evolution of the spectra of the magnetic helicity and the magnetic energy density obeys the system of equations~\cite{Boyarsky:2011uy},
%
%
\begin{align}\label{helicity_evolution2}
  \frac{\partial}{\partial\eta}\tilde{h}(\tilde{k},\eta) & =
  -\frac{2\tilde{k}^{2}}{\sigma_{c}}\tilde{h}(\tilde{k},\eta) +
  \frac{4\tilde{\Pi}}{\sigma_{c}}\tilde{\rho}_\mathrm{B}(\tilde{k},\eta),
  \notag
  \\
  \frac{\partial}{\partial\eta}\tilde{\rho}_\mathrm{B}(\tilde{k},\eta) & =
  -\frac{2\tilde{k}^{2}}{\sigma_{c}}\tilde{\rho}_\mathrm{B}(\tilde{k},\eta) +
  \frac{\tilde{\Pi}}{\sigma_{c}}\tilde{k}^{2}\tilde{h}(\tilde{k},\eta),
\end{align} 
where $\tilde{\Pi} = 2\alpha_\mathrm{em} \tilde{\mu}_{5} / \pi$ and $\tilde{\mu}_{5} = a\mu_{5}$. In Eq.~\eqref{helicity_evolution2} we assume that $\sigma_\mathrm{cond}=\sigma_cT$, where $\sigma_c\simeq 100$ in a hot QED plasma. In the following, we shall take that the initial magnetic helicity has the form,  $\tilde{h}(\tilde{k},\eta_0)=2 q \tilde{\rho}_\mathrm{B}(\tilde{k},\eta_0)/\tilde{k}$, where $0\leq q \leq 1$ is a constant parameter.
 
To complete Eq.~\eqref{helicity_evolution2} we should describe the evolution of the chiral imbalance $\mu_5=(\mu_{e\mathrm{R}} - \mu_{e\mathrm{L}})/2\neq 0$. It can be made by considering the conservation law (see, e.g., Ref.~\cite{Akamatsu:2013pjd}),
\begin{equation}\label{wellknownlaw}
  \frac{{\rm d}}{{\rm d}t}
  \left[
    (n_{e\mathrm{R}} - n_{e\mathrm{L}}) + \frac{\alpha_\mathrm{em}}{\pi}h(t)
  \right] = 0.
\end{equation}
Using Eq.~\eqref{wellknownlaw}, one gets the kinetic equation for the imbalance  $\tilde{\mu}_5 = (\xi_{e\mathrm{R}} - \xi_{e\mathrm{L}})/2$ in the form
%
%
\begin{equation}\label{mu5}
  \frac{{\rm d}\tilde{\mu}_5}{{\rm d}\eta} + \frac{6\alpha_\mathrm{em}}{\pi}
  \int \mathrm{d}\tilde{k}\frac{{\rm d}\tilde{h}(\tilde{k},\eta)}{{\rm d}\eta} =
  - \tilde{\Gamma}_f\tilde{\mu}_5,
\end{equation}
where we took into account the rate of the chirality flip $\tilde{\Gamma}_f=a\Gamma_f$ due to the nonzero $m_e\neq 0$. The explicit value of $\tilde{\Gamma}_f$ is given in Ref.~\cite{Boyarsky:2011uy}, which will be used in our analysis; cf.~Sec. \ref{sec:KOLMOGOR}.

\subsection{Monochromatic helicity spectrum: Toy model\label{sec:TOY}}

To demonstrate the possibility of the generation of the neutrino asymmetry, driven by the electroweak interaction with $e^- e^+$ plasma, we consider a maximally helical seed magnetic field, i.e. set $q=1$, and adopt monochromatic spectrum of the magnetic helicity $h(k,t)=h(t)\delta (k - k_0)$, where $k_0=r_\mathrm{D}^{-1}$, $r_\mathrm{D}=v_\mathrm{T}/\omega_p$ is the Debye length,
$\omega_p=\sqrt{4\pi\alpha_\mathrm{em} n_e/\langle E\rangle}$ is the plasma frequency. Note that $\omega_p$ coincides with the mass of a transverse plasmon in the dispersion relation $\omega=\sqrt{K^2 + \omega_p^2}$, where $\langle E\rangle\simeq 3T$ is the mean energy in a  hot ultrarelativistic plasma for which the thermal velocity $v_\mathrm{T}=1$. If one considers a non-relativistic plasma (after the positrons annihilation) with the temperature $T \ll m_e$, then $\langle E\rangle=m_e$ and $v_\mathrm{T}=\sqrt{T/m_e}$. Then we obtain from Eq.~(\ref{Fourier}) the conservation law, which is similar to Eq.~(\ref{wellknownlaw}) derived from the Adler anomaly for charged particles,
\begin{equation}\label{law}
  \frac{{\rm d}}{{\rm d}t}
  \left[
    (n_{\nu_a} - n_{\bar{\nu}_a}) - \frac{\alpha_\mathrm{ind}^a}{2\pi}h(t)
  \right]=0,
\end{equation}
where $\alpha_\mathrm{ind}^a = \left[ e^{(\nu_a)}_\mathrm{ind} \right]^2/4\pi$ is the effective electromagnetic constant and $e^{(\nu_a)}_\mathrm{ind}$ is the induced charge of a neutrino in plasma~\cite{fn4}. For a Dirac neutrino, $e^{(\nu_a)}_\mathrm{ind}$ was found in Refs.~\cite{Oraevsky:1987cu,Nieves:1993er},
\begin{equation}\label{induced}
  e_\mathrm{ind}^{(\nu_a)} =
  -\frac{G_\mathrm{F}c_\mathrm{V}^a(1 - \lambda)}{\sqrt{2}er_\mathrm{D}^2},
\end{equation}
where $\lambda=\mp 1$ is the helicity of a neutrino and the lower sign stays for a sterile particle.

Using Eq.~\eqref{induced}, one gets that
\begin{equation*}
  \left[
    e^{(\nu_a)}_\mathrm{ind}
  \right]^2 =
  7(c_\mathrm{V}^a)^2\times 10^{-14}
  \left(
    \frac{T}{m_p}
  \right)^4
\end{equation*}
in a hot plasma with $T\gg m_e$ and
$r_\mathrm{D}^{-1} = 0.075T$ given by  the electron density $n_e=0.183 T^3$. Finally, the effective coupling in Eq.~(\ref{law}) changes as the universe expands in the following way: 
%
\begin{equation}\label{alpha_nu}
  \alpha_\mathrm{ind}^a(T) =
  5.6(c_\mathrm{V}^a)^2\times 10^{-15}
  \left(
    \frac{T}{m_p}
  \right)^4.
\end{equation}

Taking for relativistic plasma, $r_\mathrm{D}=\omega_p^{-1}$, the maximum monochromatic magnetic helicity density $h=B^2/k_0=B^2/\omega_p$, from the conservation law in Eq.~(\ref{law}), one obtains the electron neutrino asymmetry at $T\gg m_e$,
\begin{equation}\label{asymmetry}
  \xi_{\nu_a}(T)=\frac{6}{\pi T^3}
  \left[
    \frac{\alpha_\mathrm{ind}^a(T)B^2}{\omega_p(T)} -   
    \frac{\alpha_\mathrm{ind}^a(T_0)B_0^2}{\omega_p(T_0)}
  \right] \approx
  -\frac{6}{\pi T^3}\frac{\alpha_\mathrm{ind}^a(T_0)B_0^2}{\omega_p(T_0)},
\end{equation}
where in the broken phase of the cooling universe at $m_e\ll T\ll T_0\ll T_\mathrm{EWPT}$ we take the zero initial asymmetry, $n_{\nu_a} - n_{\bar{\nu}_a}=\xi_{\nu_e}(T_0)T^3_0/6=\xi_{\nu_e}(T_0)=0$.

Substituting $\alpha_\mathrm{ind}^a(T)$ from Eq.~(\ref{alpha_nu}) and the seed magnetic field $B_0=0.1T_0^2$ which for the magnetic field frozen in plasma as $B=0.1T^2$ successfully obeys the limit $B< 10^{11}\,{\rm G}$ \cite{Cheng:1993kz} at the Big Bang nucleosynthesis (BBN) temperature 
$T=T_\mathrm{BBN}=0.1\,{\rm MeV}$ \cite{BBNlimit},  we obtain from Eq.~(\ref{asymmetry})
\begin{equation}\label{issue_Debye}
  \xi_{\nu_a}(T)= - 0.712(c_\mathrm{V}^a)^2 \times 10^{-15}
  \left(
    \frac{T_0}{m_p}
  \right)^4
  \left(
    \frac{T_0}{T}
  \right)^3.
\end{equation}
For the initial temperature  $T_0=1\,{\rm GeV}$ this gives at $T=\mathcal{O}({\rm MeV})$ negative asymmetry $\xi_{\nu_a}=-0.912(c_\mathrm{V}^a)^2\times 10^{-6}$. For the initial temperature $T_0=10\,{\rm GeV}$ still acceptable in the Fermi approximation for weak interactions, $T_0\ll T_\mathrm{EWPT}=100\,{\rm GeV}$, one gets a huge asymmetry $\xi_{\nu_a}= - 9.12(c_\mathrm{V}^a)^2$ that contradicts the upper limit $|\xi_{\nu_e}| < 0.07$~\cite{Dolgov:2002ab}.

Requiring that the magnitude of the asymmetry in Eq.~\eqref{issue_Debye} does not exceed the limit in Ref.~\cite{Dolgov:2002ab}, which is determined by the primordial nucleosynthesis, taking into account the equipartition of asymmetries equalization at temperatures $T = \mathcal{O}(\text{MeV})$ due to neutrino oscillations, $\xi_{\nu_e} \sim \xi_{\nu_\mu} \sim \xi_{\nu_\tau}$, we obtain the upper bound for the initial temperature, $T_0 < 5\,\text{GeV}$. Note that, at higher initial temperatures, the conservation law in Eq.~\eqref{law} should be supplemented by the evolution of the spectra of the magnetic helicity and the magnetic energy given in Eq.~\eqref{helicity_evolution2}, in which, at higher temperatures, the role of magnetic diffusion increases significantly, which was not explicitly taken into account in deriving of the result in Eq.~\eqref{issue_Debye}. Taking into account the magnetic diffusion, the generated helicity should vanish as
\begin{equation}
  h(\eta) \sim h_0(t_0)\exp
  \left[
    - \frac{2k_0^2}{\sigma_\mathrm{cond}}(t - t_0)
  \right],
\end{equation}
where $k_0^2/\sigma_\mathrm{cond} \sim T$ in the exponential.

Thus, we have demonstrated that a small-scale magnetic field, corresponding to $k_0=r_\mathrm{D}^{-1}$, with a maximal helicity, dynamo amplified in the $e^- e^+$ plasma, can feed a neutrino asymmetry through the electroweak interaction with this plasma. To refine our model, in Sec.~\ref{sec:KOLMOGOR}, we shall consider fields with a continuous  spectrum for the initial magnetic field energy density.

\subsection{Continuous Batchelor spectrum of the magnetic energy density\label{sec:KOLMOGOR}}

We shall study the neutrino asymmetry evolution basing on the continuous Batchelor spectrum of the energy density of the seed magnetic field~\cite{Dav15},
\begin{equation}\label{eq:Batchspec}
  \tilde{\rho}_\mathrm{B}(\tilde{k},\eta_0) = \mathcal{C} \tilde{k}^{\nu_\mathrm{B}},
  \quad
  \nu_\mathrm{B} = 4,
  \quad
  \mathcal{C} = \frac{(\nu_\mathrm{B}+1)\tilde{B}_{0}^{2}}{2\tilde{k}_\mathrm{max}^{\nu_\mathrm{B}+1}} =
  \frac{5\tilde{B}_{0}^{2}}{2\tilde{k}_\mathrm{max}^{5}}.
\end{equation}  
In this case $\tilde{k}_\mathrm{min}<\tilde{k}<\tilde{k}_\mathrm{max}$. As shown in Sec.~\ref{sec:TOY}, the maximal wave number $\tilde{k}_\mathrm{max}$ is constrained by the inverse Debye length in hot plasma, $\tilde{k}_\mathrm{max}<r_\mathrm{D}^{-1}/T_0 \approx 0.1$. The minimal wave number cannot be less than the inversed normalized horizon size at the given temperature, $\tilde{k}_\mathrm{min}>l_\mathrm{H}^{-1}/T_0\approx 10^{-17}$~\cite{GorRub11}, for the causal scenario to be valid. However, since $r_\mathrm{D} \ll l_\mathrm{H}$, we shall set $\tilde{k}_\mathrm{min} = 0$ for simplicity.

To proceed with the numerical analysis of Eqs.~\eqref{conformal}, \eqref{helicity_evolution2}, and~\eqref{mu5} we introduce the following new variables:
\begin{align}\label{eq:newvar}
  \mathcal{H}(\kappa,\tau)= &
  \frac{12\alpha_\mathrm{em}^{2}}{\pi^{2}}\tilde{h}(\tilde{k},\eta),
  \quad
  \mathcal{R}(\kappa,\tau) =
  \frac{24\alpha_\mathrm{em}^{2}}{\pi^{2}\tilde{k}_\mathrm{max}}\tilde{\rho}_\mathrm{B}(\tilde{k},\eta),
  \quad
  \kappa = \frac{\tilde{k}}{\tilde{k}_\mathrm{max}},
  \notag
  \\
  \mathcal{M}(\tau) = &
  \frac{2\alpha_\mathrm{em}}{\pi\tilde{k}_\mathrm{max}}\tilde{\mu}_{5}(\eta),
  \quad
  \tau=\frac{2\tilde{k}_\mathrm{max}^{2}}{\sigma_{c}}\eta,
  \quad
  \mathcal{G}=\frac{\sigma_{c}}{2\tilde{k}_\mathrm{max}^{2}}\tilde{\Gamma}_{f}.
\end{align}
where
\begin{equation}
  0<\kappa<1,
  \quad
  \tau > \tau_0 =
  1.44\times10^{13}
  \left(
    \frac{\tilde{k}_\mathrm{max}}{0.1}
  \right)^2
  \left(
    \frac{T_0}{10\,\text{GeV}}
  \right)^{-1}.
\end{equation}
Using Eq.~\eqref{eq:Batchspec}, we get that the initial values of the functions $\mathcal{H}(\kappa,\tau)$, $\mathcal{M}(\tau)$, and $\mathcal{R}(\kappa,\tau)$ are
\begin{align}
  \mathcal{H}(\kappa,\tau_0) = & q \mathcal{R}_{0} \kappa^{\nu_\mathrm{B}-1},
  \quad
  \mathcal{M}(\tau_{0})=4.66\times10^{-2}\tilde{\mu}_{5}(\eta_0)
  \left(
    \frac{\tilde{k}_\mathrm{max}}{0.1}
  \right)^{-1}
  \notag
  \\
  \mathcal{R}(\kappa,\tau_{0}) = & \mathcal{R}_{0}\kappa^{\nu_\mathrm{B}},
  \quad
  \mathcal{R}_{0} =
  \frac{12\alpha_\mathrm{em}^{2}\tilde{B}_{0}^{2}(\nu_\mathrm{B}+1)}
  {\pi^{2}\tilde{k}_\mathrm{max}^2} =
  3.24\times 10^{-4}
  \left(
    \frac{\tilde{k}_\mathrm{max}}{0.1}
  \right)^{-2},
\end{align}
where we assume that the initial helicity spectrum reads $\tilde{h}(\tilde{k},\eta_0) = 2q \tilde{\rho}_\mathrm{B}(\tilde{k},\eta_{0})/\tilde{k}$, with $0\leq q \leq 1$, and $\tilde{B}_{0}= a^2 B_{0} = 0.1$; cf. Sec.~\ref{sec:TOY}.

The kinetic Eqs.~\eqref{helicity_evolution2} and~\eqref{mu5} can be rewritten in the form,
\begin{align}\label{eq:HRM}
  \frac{\partial\mathcal{H}}{\partial\tau} = &
  -\kappa^{2}\mathcal{H}+\mathcal{M}\mathcal{R},
  \notag
  \\
  \frac{\partial\mathcal{R}}{\partial\tau} = &
  -\kappa^{2}\mathcal{R}+\kappa^{2}\mathcal{M}\mathcal{H},
  \notag
  \\
  \frac{\mathrm{d}\mathcal{M}}{\mathrm{d}\tau}	= &
  \int_{0}^{1}d\kappa\left(\kappa^{2}\mathcal{H}-\mathcal{M}\mathcal{R}\right) -
  \mathcal{G}\mathcal{M},
\end{align}
where
\begin{equation}\label{eq:flipratenorm}
  \mathcal{G} = 3.57 \times 10^{-37} \tau^2
  \left(
    \frac{\tilde{k}_\mathrm{max}}{0.1}
  \right)^{-6}
\end{equation}
In Eqs.~\eqref{eq:HRM} and~\eqref{eq:flipratenorm}, we use the new variables in Eq.~\eqref{eq:newvar}.

The neutrino asymmetry, resulting from the integration of the master Eq.~(\ref{conformal}),
\begin{align}\label{asymmetry_final}
  \xi_{\nu_e}(\tau) = &
  \xi_{\nu_e}(\tau_{0})
  \notag
  \\
  & +
  \frac{A^{2}M_{0}^{4}\tilde{k}_\mathrm{max}^{13}}{\alpha_\mathrm{em}^{2}\sigma_{c}^{4}}
  \int_{0}^{1}\kappa^{4}d\kappa
  \left[
    \frac{\mathcal{H}(\kappa,\tau)}{\tau^{4}} -
    \frac{\mathcal{H}(\kappa,\tau_{0})}{\tau_{0}^{4}} +
    2\int_{\tau_{0}}^{\tau}d\tau'\frac{\mathcal{H}(\kappa,\tau')}{\tau^{\prime5}}
  \right],
\end{align}
is determined, in general, by the solution of the self-consistent Eq.~(\ref{eq:HRM}). In Eq.~\eqref{asymmetry_final}, we express $\xi_{\nu_e}(\eta)$ in terms of the new variables in Eq.~\eqref{eq:newvar}.

\begin{figure}
  \centering
  \subfigure[]
  {\label{1a}
  \includegraphics[scale=.19]{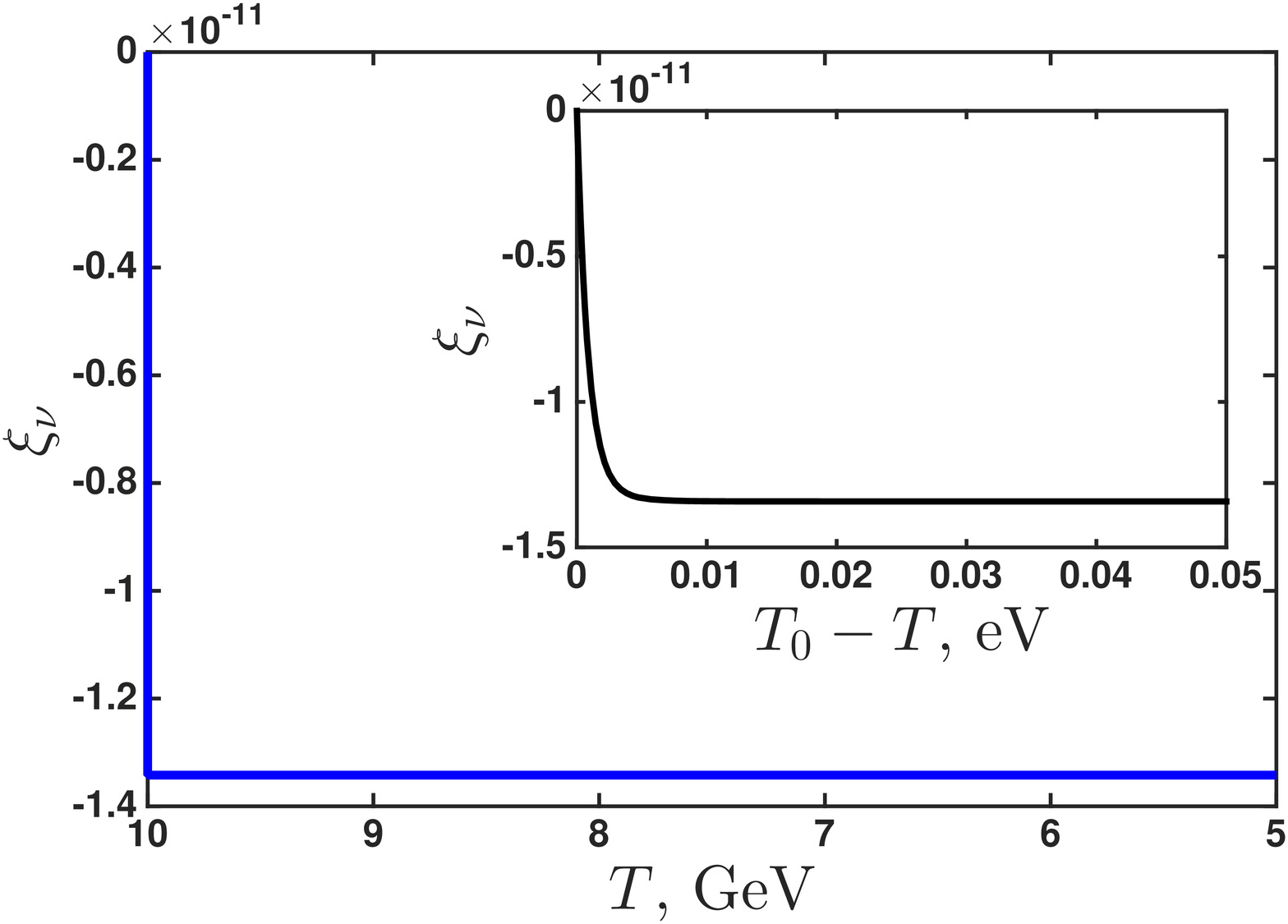}}
  \hskip-.7cm
  \subfigure[]
  {\label{1b}
  \includegraphics[scale=.19]{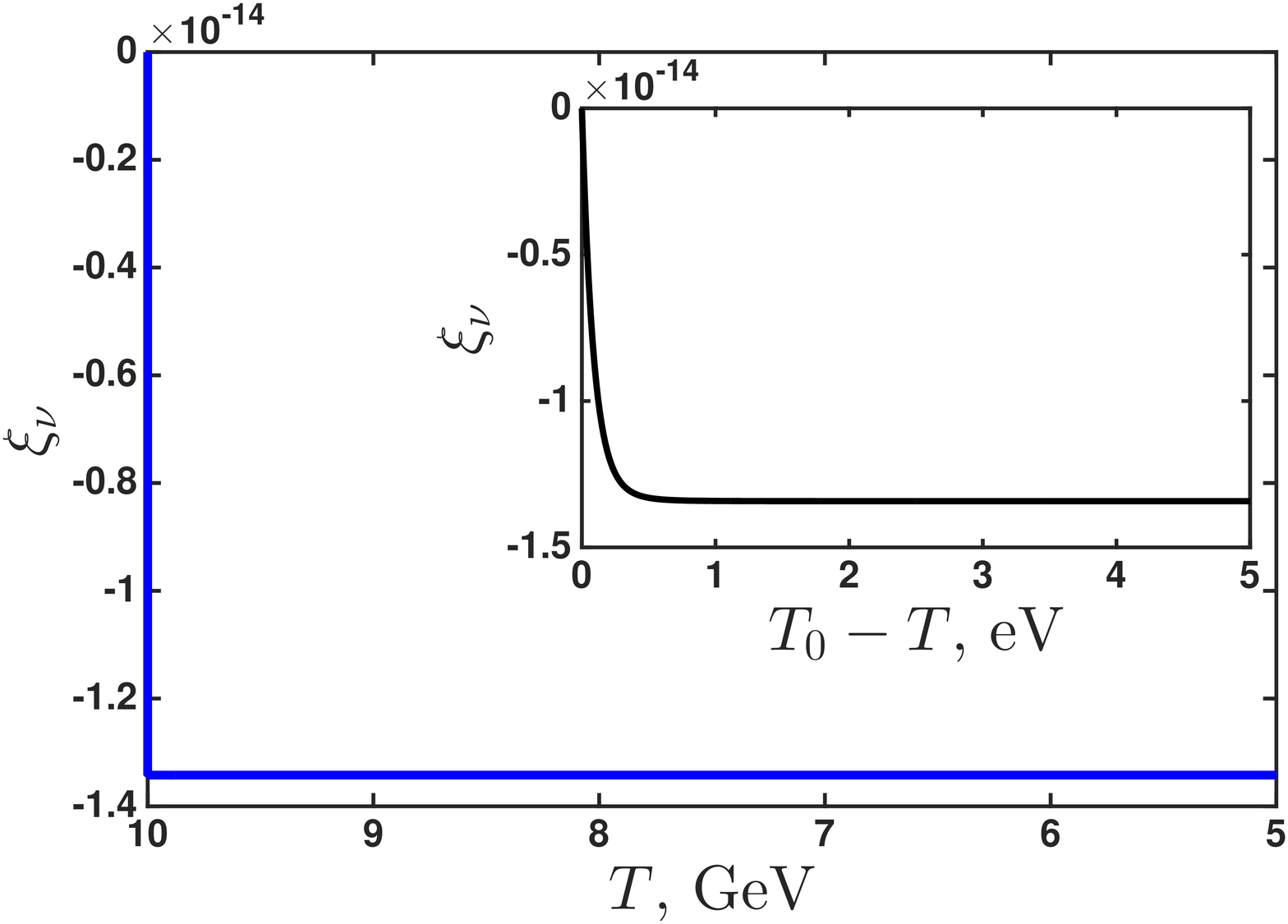}}
  \\
  \subfigure[]
  {\label{1c}
  \includegraphics[scale=.19]{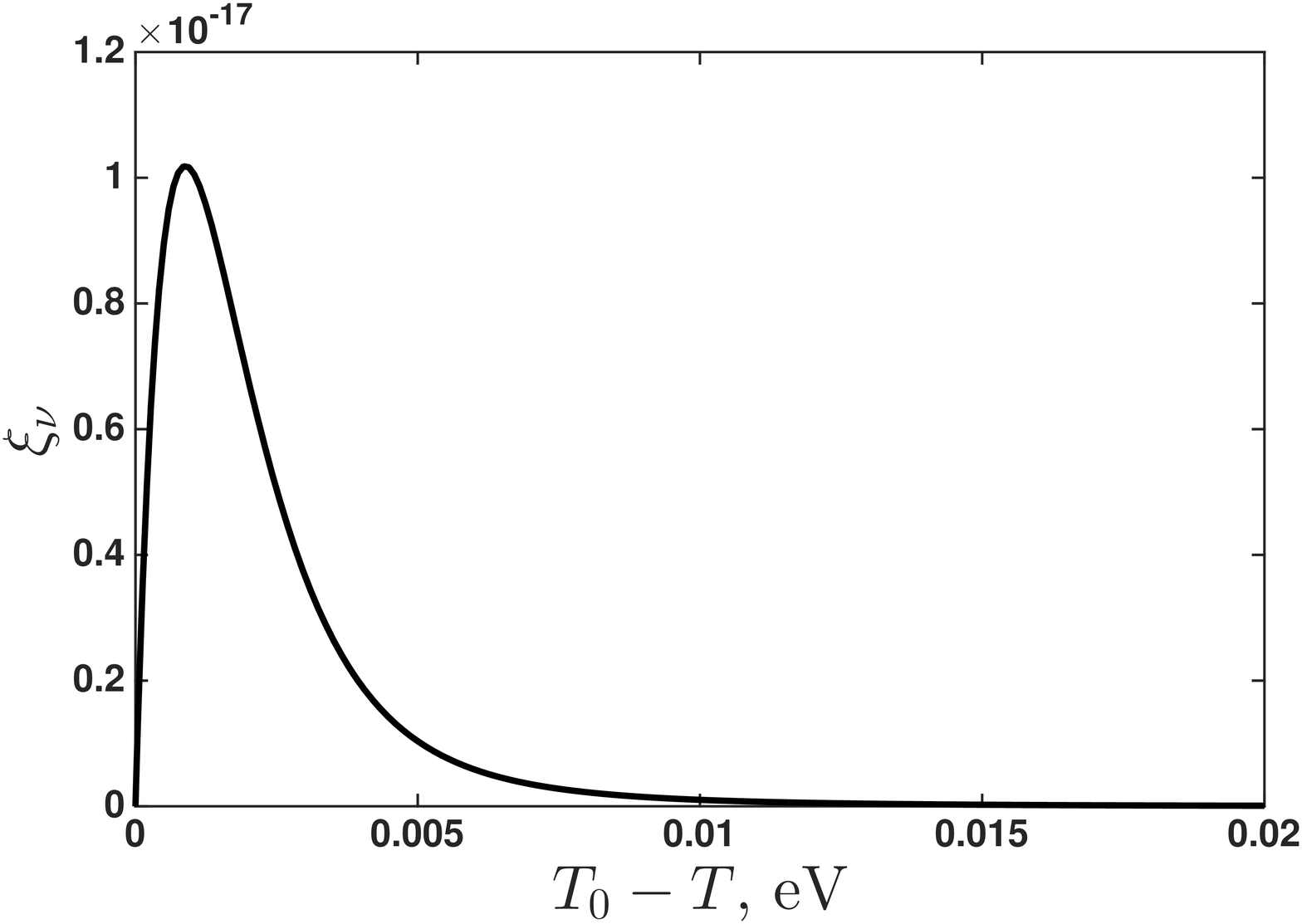}}
  \hskip-.7cm
  \subfigure[]
  {\label{1d}
  \includegraphics[scale=.19]{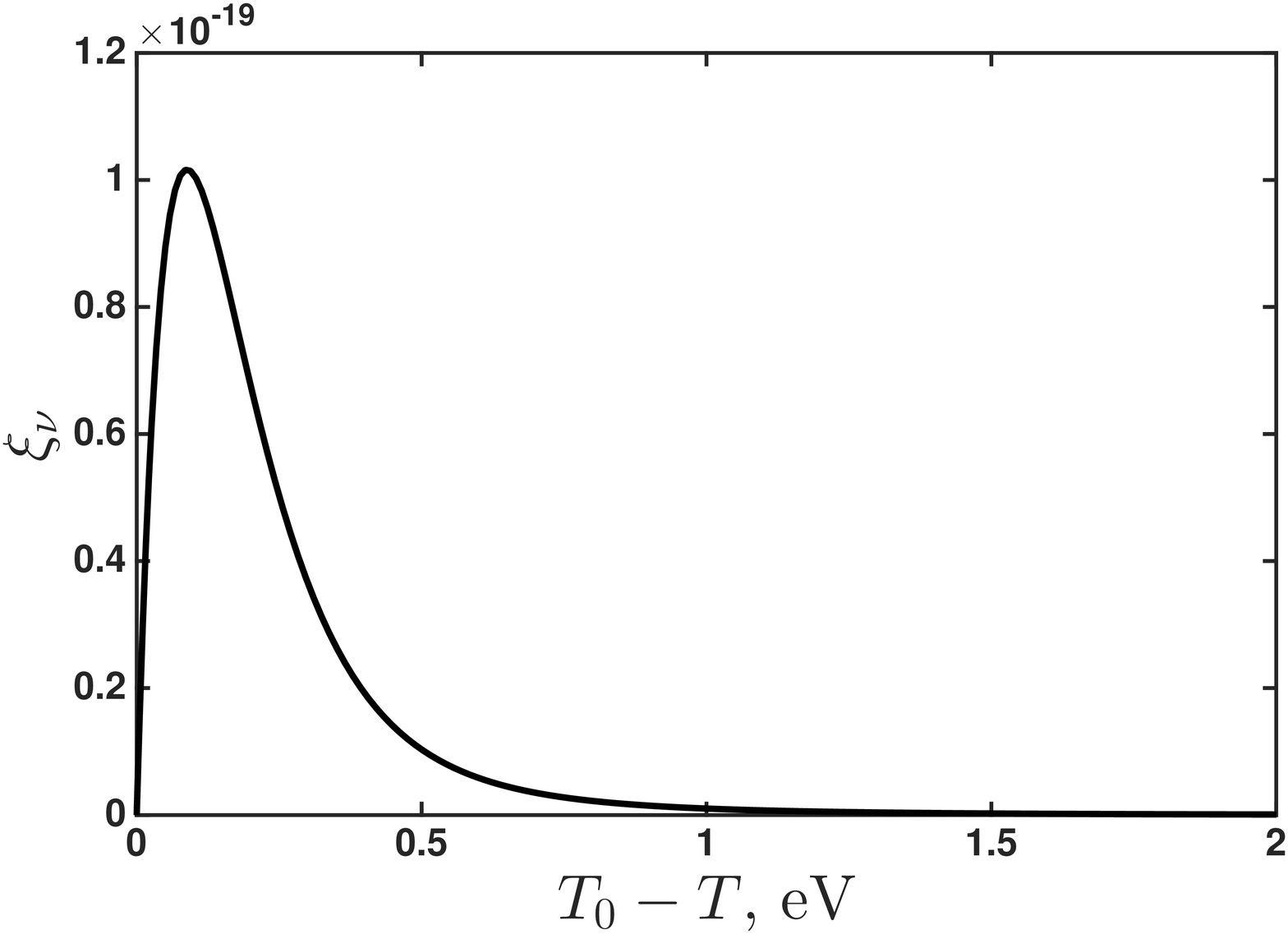}}
  \protect
  \caption{\label{fig:xievol}
    The behavior of $\xi_\nu$ versus $T$
    for different $\tilde{k}_\mathrm{max}$ and $q$, assuming that the initial asymmetry
    is absent, $\xi_{\nu}(T_0)=0$.
    The neutrino asymmetry in panels~(a) and~(b) is represented by
    blue lines to distinguish them from the vertical axes.
    (a) $\tilde{k}_\mathrm{max} = 10^{-1}$ and $q=1$,
    (b) $\tilde{k}_\mathrm{max} = 10^{-2}$ and $q=1$,
    (c) $\tilde{k}_\mathrm{max} = 10^{-1}$ and $q=0$,
    (d) $\tilde{k}_\mathrm{max} = 10^{-2}$ and $q=0$.
    The insets in panels~(a) and~(b) show $\xi_\nu$
    at small evolution times.
  }
\end{figure}

We solve Eq.~\eqref{eq:HRM} numerically assuming that $\tilde{\mu}_{5}(\eta_{0}) = 4\times10^{-5}$. Analogous initial chiral imbalance was considered in Ref.~\cite{Boyarsky:2011uy}. The resulting neutrino asymmetry $\xi_\nu \equiv \xi_{\nu_e}$ is shown in Fig.~\ref{fig:xievol}, where we take that $\xi_{\nu}(\tau_{0}) = 0$. In Fig.~\ref{fig:xievol}, we start with the initial temperature $T_0 = 10\,\text{GeV}$ and plot $\xi_{\nu}(\tau)$ for different $\tilde{k}_\mathrm{max}$ and $q$. The Fermi approximation is valid for these temperatures since $T_0 \ll M_\mathrm{W} \sim 10^2\,\text{GeV}$. One can see in Fig.~\ref{fig:xievol} that $\xi_\nu$ reaches the asymptotic value very rapidly. Therefore, we restrict ourselves by discussing $T_0 < T< 5\,\text{GeV}$.

Comparing Figs.~\ref{1a} and~\ref{1b} with Figs.~\ref{1c} and~\ref{1d}, one can see that the asymptotic neutrino asymmetry $\xi_{\nu}^{(\infty)}$ at $T\ll T_0$, which turns out to be negative, is purely defined by the initial helicity of the magnetic field. Using Eq.~\eqref{asymmetry_final}, one obtains that the value of $\xi_{\nu}^{(\infty)}$ reads,
\begin{equation}\label{eq:asympt}
  \xi_{\nu}^{(\infty)} = - \frac{15}{32 \pi^2}
  q A^2 \tilde{B}_0^2 T_0^4 \tilde{k}_\mathrm{max}^3.
\end{equation}
Taking $q=1$, $A = 5.29\times10^{-5}\,\text{GeV}^{-2}$, $\tilde{B}_0=0.1$, $T_0 = 10\,\text{GeV}$, and $\tilde{k}_\mathrm{max} = 10^{-1}$, one gets that $\xi_{\nu}^{(\infty)} \approx - 1.3 \times 10^{-11}$, that is in good agreement with Fig.~\ref{1a}. Moreover, setting $q=0$ in Eq.~\eqref{eq:asympt}, which corresponds to a nonhelical seed field, one gets that $\xi_{\nu}^{(\infty)} = 0$. This result can be also seen in Fig.~\ref{1c}.

In Figs.~\ref{1a} and~\ref{1c}, we take $\tilde{k}_\mathrm{max} = 10^{-1}$, which was mentioned above to be the maximal possible wave number in a hot relativistic plasma. To examine the dependence of $\xi_{\nu}^{(\infty)}$ on the length scale of the magnetic field, in Figs.~\ref{1b} and~\ref{1d}, we plot the neutrino asymmetry for the smaller $\tilde{k}_\mathrm{max} = 10^{-2}$. One can see in Fig.~\ref{1b} that  $\xi_{\nu}^{(\infty)}$ is three orders of magnitude smaller than that in Fig.~\ref{1a}, that is again in good agreement with Eq.~\eqref{eq:asympt}.



\section{Conclusion\label{Conclusions}}

In this work we have derived the new kinetic equations for ultrarelativistic neutrinos in a hot plasma of the early Universe. These equations account for both the Berry curvature terms and the electroweak interaction of neutrinos with background fermions. Basing on these kinetic equations we have derived a new anomaly in plasma after EWPT, related to the nonconservation of the lepton current because of the Berry curvature, and recovered the well-known abelian anomaly for lepton currents in hypercharge fields before EWPT, without using the Feynman diagrams technique. Then we have applied the obtained anomaly to generate the neutrino asymmetries in the broken phase of the early Universe.

Using our results, in Sec.~\ref{afterEWPT}, one can conclude that the new neutrino anomaly, describing the nonconservation of the neutrino current in Eq.~(\ref{main}) at temperatures much below EWPT, $T\ll T_\mathrm{EWPT}$, results in the neutrino asymmetry in Eq.~\eqref{asymmetry_final} for a realistic continuous spectrum of the magnetic energy density. In the case of the abelian anomaly in Eq.~(\ref{abelian}), derived here independently in Eq.~(\ref{main2}) through the generalization  of the Boltzmann equation for neutrinos in Eq.~(\ref{RKEbefore}) with help of the Berry curvature in Eq.~(\ref{curvature}), appears to be more efficient in the generation of the neutrino asymmetry. Considering the neutrino asymmetry before EWPT at $T> T_\mathrm{EWPT}$, e.g., in Ref.~\cite{Semikoz:2013xkc}, one gets that $\xi_{\nu_{e\mathrm{L}}}$ can be as high as $10^{-10}$ for some hypermagnetic spatial scales $\tilde{k}_0^{-1}$ at $T_\mathrm{EWPT}\simeq 100\,{\rm GeV}$. This value is  close to the observable BAU $\sim 10^{-10}$~\cite{GorRub11}.

Nevertheless, our results are important since, for the first time we have shown that a neutrino asymmetry can be generated due to the combination of two factors: the additional Berry curvature terms in the Boltzmann kinetic equation in the momentum space as a part of the full phase volume for the neutrino distribution function, and the electroweak interaction of neutrinos with background fermions. The predicted effect can happen even at zero seed neutrino asymmetry owing to the nonconservation of the neutrino lepton current, related to the nonconservation of the neutrino four-current in matter permeated by an external magnetic (hypermagnetic) field.

In a common opinion, after the neutrino decoupling at $T< T_\mathrm{dec}\sim (2-3)\,{\rm MeV}$ in the non-relativistic plasma $T\ll m_e$ during the radiation dominated era (with the red shift $z>10^4$) the neutrino free streaming means that the neutrino asymmetries freeze-out at the final level of the neutrino decoupling. However, the neutrino current, in fact, still is not conserved, $\partial_{\mu}j_{\nu_a}^{\mu}\neq 0$, because of the Berry curvature in the momentum space through the Boltzmann equation in Eq.~(\ref{Boltzmann2}) in the Vlasov approximation, i.e. without the collision integrals. We have studied this case as well and found that the growth of the neutrino asymmetry is nonzero but quite small even compared to that in Eq.~(\ref{asymmetry_final}). Of course, the electric conductivity in this situation becomes different: $\sigma_\mathrm{cond}\sim T^{3/2}$ instead of $\sigma_\mathrm{cond}\sim T$  in Eq.~(\ref{asymmetry_final}) based on the magnetic helicity evolution in Eq. (\ref{helicity_evolution2}).

We conclude that a large value of the neutrino asymmetry close to the baryon asymmetry $\sim 10^{-10}$ can be acquired mostly before EWPT through the quantum (abelian) anomalies in Eqs.~(\ref{main2}), (\ref{main3}) driven by the external hypermagnetic field (see, e.g., Ref.~\cite{Semikoz:2013xkc}).  In the present work we have derived these anomalies by a new method modifying the Boltzmann Eq.~(\ref{RKEbefore}) through the Berry curvature. Nevertheless, as results from Fig.~\ref{fig:xievol}, the neutrino asymmetry generated after EWPT in our model is only $\sim 8$ times less than the observed BAU.

\section*{Acknowledgments}
One of the authors (MD) is thankful to the Competitiveness Improvement Program at the Tomsk State University and RFBR (research project No.~15-02-00293) for a partial support.

\appendix

\section{Conservation of the neutrino current in elactic and inelastic processes\label{sec:CONTEQ}}

In this appendix we shall demonstrate that the neutrino current is conserved in both inelastic processes and elastic scattering off backfround electrons. For this purpose we shall consider the primordial plasma above neutrino decoupling at $T>T_\mathrm{dec}= (2 - 3)\,{\rm MeV}$.

Firstly we mention that, if the Berry curvature is not taken into account, the neutrino current conservation in Eq.~\eqref{current} has the form of the continuity equation which results from the Boltzmann kinetic equations given in Eqs.~\eqref{RKEbefore} and~\eqref{Boltzmann} integrated over neutrino momenta $\mathrm{d}^3k$. The contribution of the collision integrals in the rhs of these equations is vanishing: $\smallint \mathrm{d}^3k J^{(\nu_a)}_\mathrm{coll}[f_{\bf k}]=0$, where $f_{\bf k}$ is the neutrino distribution function.

Let us demonstrate that the neutrino current is conserved in a hot background $e^+e^-$ plasma at the thermal equilibrium with neutrinos and antineutrinos considering first the inelastic process $\nu + \bar{\nu}\leftrightarrow e^+ + e^-$. The collision integral for this process was found in Ref.~\cite[Eq.~(A9)]{Bru85},
\begin{equation}\label{eq:Jcollie}
  J_\mathrm{coll}^{(\nu)}[f_{\bf q}] =
  \int \frac{\mathrm{d}^3 q'}{(2\pi)^3}
  \left[
    R_{{\rm TP}}^{(p)}({\bf q}, {\bf q}')
    (1 - f^{(\nu)}({\bf q})(1 - f^{(\bar{\nu})}({\bf q}') -
    R_{{\rm TP}}^{(a)}({\bf q}, {\bf q}')
    f^{(\nu)}({\bf q})f^{(\bar{\nu})}({\bf q}')
  \right],
\end{equation}
where $f^{(\bar{\nu})}({\bf q}')$ is the antineutrino distribution function. The factor $R_{{\rm TP}}^{(p)}({\bf q}, {\bf q}')$, which corresponds to the neutrino pair production, has the form~\cite[Eq.~(C6)]{Bru85},
\begin{equation}\label{eq:Rp}
  R_{{\rm TP}}^{(p)} =
  \int \frac{\mathrm{d}^3p_e}{(2\pi)^3}
  \int \frac{\mathrm{d}^3p_{e^+}}{(2\pi)^3} 2f^{(e)}(E_{e^-}) 2f^{(\bar{e})}(E_{e^+})
  r^{(p)}(p_e + p'_{e^+}\to q + q'),
\end{equation}
whereas the factor $R_{{\rm TP}}^{(a)}({\bf q}, {\bf q}')$, corresponding to the neutrino pair annihilation $\nu\bar{\nu}\to e^+e^-$ is given by the following expression~\cite[Eq.~(C6)]{Bru85}:
\begin{equation}\label{eq:Ra}
  R_{{\rm TP}}^{(a)}=
  \int \frac{\mathrm{d}^3p_e}{(2\pi)^3}
  \int \frac{\mathrm{d}^3p_{e^{+}}}{(2\pi)^3}
  [1 -f^{(e)}(E_{e^-})][1 - f^{(\bar{e})}(E_{e^+})]
  r^a(q + q^{'}\to p_e + p_{e^+}),
\end{equation}
where $E_{e^\mp}$ are the energies of electrons and positrons.

The rates of the annihilation $r^{(a)}(q + q' \to p_e + p_{e^+})$ and the production $r^{(p)}(p_e + p_{e^+}\to q + q')$ are related by the reciprocity condition~\cite[Eq.~(C3)]{Bru85}, $r^{a}=4r^p$. Thus it is sufficient to define only, e.g., $r^p$~\cite[Eq.~(C52)]{Bru85},
\begin{multline}\label{eq:rp}
  r^p(p_e + p_{e^+}\to q + q') =
  \frac{G_\mathrm{F}^2}{2\omega\omega^{'}E_eE_{e^+}}(2\pi)^4
  \delta^{(4)}(q + q' - p_e - p_{e^+})
  \\
  \times
  \left[
    (C_\mathrm{V} + C_\mathrm{A})^2(p_{e^+}\cdot q)(p_e\cdot q') +
    (C_\mathrm{V} -C_\mathrm{A})^2(p_{e^+}\cdot q')(p_e\cdot q) +
    m_e^2(C_\mathrm{V}^2 - C_\mathrm{A}^2)(q\cdot q')
  \right].
\end{multline}
where $C_\mathrm{V,A}$ are the vector and axial-vector constants of the standard model, and $\omega$ and $\omega'$ are the neutrino and antineutrino energies.

Finally, using Eqs.~\eqref{eq:Ra}-\eqref{eq:rp}, one can rewrite the full Boltzmann integral for inelastic processes $\nu\bar{\nu} \leftrightarrow e\bar{e}$ in Eq.~\eqref{eq:Jcollie} in the following form:
\begin{align}\label{eq:Jcolliefin}
  J_\mathrm{coll}^{(\nu)}[f_{\bf q}] = &
  \frac{2G_\mathrm{F}^2}{\omega}
  \int \frac{\mathrm{d}^3 q'}{(2\pi)^3\omega'}
  \int \frac{\mathrm{d}^3p_e}{(2\pi)^3E_e}
  \int \frac{\mathrm{d}^3p_{e^{+}}}{(2\pi)^3E_{e^+}}(2\pi)^4
  \delta^{(4)}(q + q' - p_e - p_{e^+})
  \notag
  \\
  & \times
  \left[
    (C_\mathrm{V} + C_\mathrm{A})^2(p_{e^+}\cdot q)(p_e\cdot q') +
    (C_\mathrm{V} -C_\mathrm{A})^2(p_{e^+}\cdot q')(p_e\cdot q) +
    m_e^2(C_\mathrm{V}^2 - C_\mathrm{A}^2)(q\cdot q')
  \right]
  \notag
  \\
  & \times
  \big\{
    [1 - f^{(\nu)}({\bf q})][1 - f^{(\bar{\nu})}({\bf q}')]
    f^{(e)}(E_{e^-})f^{(\bar{e})}(E_{e^+})
    \notag
    \\
    & -
    f^{(\nu)}({\bf q})f^{(\bar{\nu})}({\bf q}')
    [1 -f^{(e)}(E_{e^-})][1 - f^{(\bar{e})}(E_{e^+})]
  \big\}.
\end{align}
The last multiplier in the integrand in Eq.~\eqref{eq:Jcolliefin} is called the Boltzmann's brace. It combines the Fermi distribution functions for particles and antiparticles,
$f^{(a,\bar{a})}(E_{a,\bar{a}})= [1 + \exp ((E_{a,\bar{a}} \mp \mu_a)/T)]^{-1}$, with the lower sign staying for antiparticles.

We recall that we consider the primordial plasma with temperatures above the neutrino decoupling. In this case the Boltzmann brace equals to zero since $\exp [(\omega + \omega^{'})/T] - \exp[(E_{e^-} + E_{e^+})/T] = 0$ owing to the energy conservation law, accounted for by the energy-momentum conservation delta function, and the equilibrium, $T=T_{\nu}=T_{e}$. Thus inelastic processes $\nu\bar{\nu}\leftrightarrow e\bar{e}$ do not violate the neutrino current conservation in Eq.~\eqref{current}.

Now let us consider the neutrino elastic scattering (NES) off background electrons $\nu e\to \nu e$. In this situation, the Boltzmann collision integral takes the form~\cite[Eq.~(A6)]{Bru85},
\begin{align}\label{eq:Jcollel}
  J_\mathrm{coll}^{(\nu)}[f_{\bf q}] = &
  \int \frac{\mathrm{d}^3q'}{(2\pi)^3}
  \Big\{
    R_{{\rm NES}}^{(\mathrm{in})}({\bf q}, {\bf q'})
    [1 - f^{(\nu)}({\bf q})] f^{(\nu)}({\bf q}')
    \notag
    \\
    & -
    R_{{\rm NES}}^{(\mathrm{out})}({\bf q}, {\bf q}')
    f^{(\nu)}({\bf q})[1 - f^{(\nu)}({\bf q}')]
  \Big\},
\end{align}
where the factors $R_{{\rm NES}}^{(\mathrm{in})}({\bf q}, {\bf q'})$ and $R_{{\rm NES}}^{(\mathrm{out})}({\bf q}, {\bf q'})$ have the form,
\begin{align}\label{eq:Rinout}
  R_{{\rm NES}}^{(\mathrm{in})} = &
  \int \frac{\mathrm{d}^3p_e}{(2\pi)^3}
  \int \frac{\mathrm{d}^3 p'_e}{(2\pi)^3}[1 - f^{(e)}({\bf p})]2f^{(e)}({\bf p}')
  r^{(\mathrm{in})}(p'_e + q'\to p_e + q),
  \notag
  \\
  R_{{\rm NES}}^{(\mathrm{out})} = &
  \int \frac{\mathrm{d}^3p_e}{(2\pi)^3}
  \int \frac{\mathrm{d}^3p'_e}{(2\pi)^3}[1 - f^{(e)}({\bf p}^{'})]2f^{(e)}({\bf p})
  r^{(\mathrm{out})}( p_e + q\to p'_e + q').
\end{align}
Here the rates $r^{(\mathrm{in})}$ and $r^{(\mathrm{out})}$ are equal due to the reciprocity condition~\cite[Eq.~(C6)]{Bru85}: $r^{(\mathrm{in})} = r^{(\mathrm{out})} = r$. The value of $r$ for NES is given in Ref.~\cite[Eq.~(C49)]{Bru85},
\begin{multline}\label{eq:r}
  r = \frac{G_\mathrm{F}^2}{\omega\omega' E_eE_e^{'}}
  (2\pi)^4 \delta^{(4)}(q + p_e - q' - p'_e)
  \\
  \times
  \left[
    (C_\mathrm{V} + C_\mathrm{A})^2 (p_e\cdot q)(p'_e\cdot q') +
    (C_\mathrm{V} - C_\mathrm{A})^2 (p'_e\cdot q)(p_e\cdot q') -
    m_e^2(C_\mathrm{V}^2 - C_\mathrm{A}^2) (q \cdot q')
  \right].
\end{multline}
Note that $r$ in Eq.~\eqref{eq:r} is symmetric under the interchange of the momenta $p_e\leftrightarrow p'_e$ and $q\leftrightarrow q'$.

Using Eqs.~\eqref{eq:Rinout} and~\eqref{eq:r}, we get the Boltzmann collision integral for NES,
\begin{align}\label{eq:Jcollelfin}
  J_\mathrm{coll}^{(\nu)}[f_{\bf q}] = &
  \frac{2G_\mathrm{F}^2}{\omega}
  \int \frac{\mathrm{d}^3q'}{(2\pi)^3\omega^{'}}
  \int \frac{\mathrm{d}^3p_e}{(2\pi)^3E_e}
  \int \frac{\mathrm{d}^3p'_{e}}{(2\pi)^3E'_{e}}
  (2\pi)^4\delta^{(4)}(q + p_e - p'_e -q' )
  \notag
  \\
  & \times
  \left[
    (C_\mathrm{V} + C_\mathrm{A})^2 (p_e\cdot q)(p'_e\cdot q') +
    (C_\mathrm{V} - C_\mathrm{A})^2 (p'_e\cdot q)(p_e\cdot q') -
    m_e^2(C_\mathrm{V}^2 - C_\mathrm{A}^2) (q \cdot q')
  \right]
  \notag
  \\
  & \times    
  \big\{
    [1 - f^{(\nu)}({\bf q})][1 - f^{(e)}({\bf p})]
    f^{(\nu)}({\bf q}')f^{(e)}({\bf p}') 
    \notag
    \\
    & -
    [1 - f^{(\nu)}({\bf q}')][1 - f^{(e)}({\bf p}')]
    f^{(\nu)}({\bf q})f^{(e)}({\bf p})
  \big\}. 
\end{align}
Then we use the property of the Boltzmann brace in the last multiplier in the integrand in Eq.~\eqref{eq:Jcollelfin}, that it changes the sign under the interchange of the inner variables, $p_e\leftrightarrow p'_e$ and $q\leftrightarrow q'$. This property remains valid for arbitrary distribution functions including those beyond the equilibrium. Therefore, if one integrates Eq.~\eqref{eq:Jcollelfin} over $\mathrm{d}^3q$, the 12-fold integral vanishes giving one $\smallint \mathrm{d}^3q J_\mathrm{coll}^{(\nu)} = 0$. We should recall that, to get the continuity equation from the Boltzmann kinetic equation, one should integrate it over the momentum. Thus, NES does not contribute to Eq.~\eqref{current} either. Note that this fact also follows from the results of Ref.~\cite{Semikoz:1987cv}, where the $\nu e$ elastic scattering was studied.

Finally, we have shown that the conservation of the neutrino current in Eq.~\eqref{current} in equilibrium primordial plasma is valid when both elastic and inelastic processes are taken into account. Thus, the nonconservation of the current can be only due to the consideration of the Berry curvature effects as suggested in Sec.~\ref{RKE}.


\begin{thebibliography}{100}
\bibitem{fn4}
  In Ref.~\cite{Oraevsky:1987cu}, the induced charge for the Dirac neutrino
  was found as
  $e_\mathrm{ind}^{(\nu_a)} =
  -G_\mathrm{F}|e|c_\mathrm{V}/2\pi\sqrt{2}\alpha_\mathrm{em} r_\mathrm{D}^2$,
  which coincides with Eq.~(\ref{induced}) here taken from Eq.~(5.16)
  in Ref.~\cite{Nieves:1993er}.
\end{thebibliography}
\end{document}